\documentclass[aip,graphicx]{revtex4-1}
\usepackage{graphicx}
\usepackage{soul}
\usepackage{amsmath}
\usepackage{amsfonts}
\usepackage{amssymb}
\usepackage{multirow}
\usepackage{natbib}
\usepackage{soul,xcolor}

\setstcolor{red}

\newcommand{\Ket}[1]{{\left|{#1}\right\rangle}}

\draft 

\begin{document}


\title{InGaAs/AlInGaAs THz Quantum Cascade Lasers operating up to 195 K in strong magnetic field} 


\author{Federico Valmorra}
\email[]{valmorrf@ethz.ch}
\author{Giacomo Scalari}
\author{Keita Ohtani}
\author{Mattias Beck}
\author{J\'er\^ome Faist}
\affiliation{Institute for Quantum Electronics, ETH Zurich, Auguste-Piccard-Hof 1, CH-8093 Zurich, Switzerland}



\date{\today}

\begin{abstract}
Terahertz quantum cascade lasers based on InGaAs wells and quaternary AlInGaAs barriers were measured in magnetic field. This study was carried out on a four quantum well active region design with photon energy of 14.3 meV processed both with Au and Cu waveguides. The heterostructure operates up to 148 K at B=0 T and in a Cu waveguide. The complete magneto-spectroscopic study allowed the comparison of emission and transport data. Increasing the magnetic field, the low effective mass of the InGaAs wells allowed us to reach the very strong confinement regime. At B=12 T, where the cyclotron transition is almost resonant with the LO-phonon, we recorded a maximum operating temperature of 195 K for the devices with Cu waveguide. Additional lasing at 5.9 meV was detected for magnetic fields between 7.3 and 7.7 T.
\end{abstract}

\pacs{}

\maketitle 


The terahertz (THz) spectral range is regarded with ever increasing interest for sensing, imaging and spectroscopy applications and quantum cascade lasers (QCLs) represent a primary semiconductor-based, electrically pumped source that can cover its range.\cite{Kohler2002,Williams2007,ScalariLPR2009,Fathololoumi2012,Jung2014,Lu2014} THz QCLs in the GaAs/AlGaAs material system saw big improvements both in frequency coverage and operating temperature since their first demonstration, but room temperature operation is still missing. 
In order to increase the gain, and therefore the operating temperature of THz-QCLs, InGaAs based active regions were investigated, first in combination with InAlAs barriers reaching 122 K\cite{Fischer2010} and, more recently, with GaAsSb reaching 142 K\cite{Deutsch2012}. Such expectations stem from the beneficial lower electron effective masses of all these materials with respect to the GaAs/AlGaAs material system, thus allowing higher oscillator strength and gain. On the other hand, one of the most relevant limiting factors, especially for InGaAs/GaAsSb QCLs, is the interface asymmetry that causes strong elastic electron scattering\cite{Deutsch2013}. Furthermore both ternary compounds are lattice matched to InP for a single stoichiometry that has a high conduction band offset that results in very thin barriers, very sensitive to inherent thickness fluctuations from the growth process.  An alternative option comes from QCLs based on quaternary AlInGaAs barrier material.\cite{Olego1982, Alavi1983} InGaAs/AlInGaAs QCLs were recently demonstrated, reporting a maximum power $P_{max}=35$mW at 3.8~THz at 10~K and a maximum operating temperature $T_{max}=130$~K.\cite{Ohtani2013} This material system maintains the beneficial low electron effective masses while presenting a more symmetric interface and a lower, more suitable conduction band offset that can be adjusted by the composition for the quaternary barrier material.

The application of magnetic field along the growth axis is a very useful tool to investigate the QCL operation and to identify the different scattering mechanisms\cite{Leuliet2006,Fischer2010, PereLaperne2007} and can as well enhance the gain because of the selective closing of electronic channels\cite{Scalari2010} allowing a higher operating temperature. Such kind of study was performed for the different material systems where THz QCLs have been realised. The GaAs/AlGaAs QCL reached 225 K at 19.3 T\cite{Wade2009}, while the InGaAs/GaAsSb QCL showed laser action up to 190 K at 11 T\cite{Maero2013}. The low effective mass of InGaAs allows and fosters these studies that can be performed with lab-size superconducting magnets, contrary to the high-field facilities required for GaAs-based devices. In this letter we present a study in magnetic field for InGaAs/AlInGaAs THz QCLs. We will also show that, thanks to magnetic field enhanced gain, the QCL processed in a Cu-Cu waveguide reaches a maximum temperature of 195 K for an applied magnetic field of 12 T. We also present comparative measurements for the same active region as in Ref.\cite{Ohtani2013} processed once in Au-Au waveguide and then in Cu-Cu waveguide.

\begin{figure}[htbp]
\centering
\includegraphics[width=.9\linewidth]{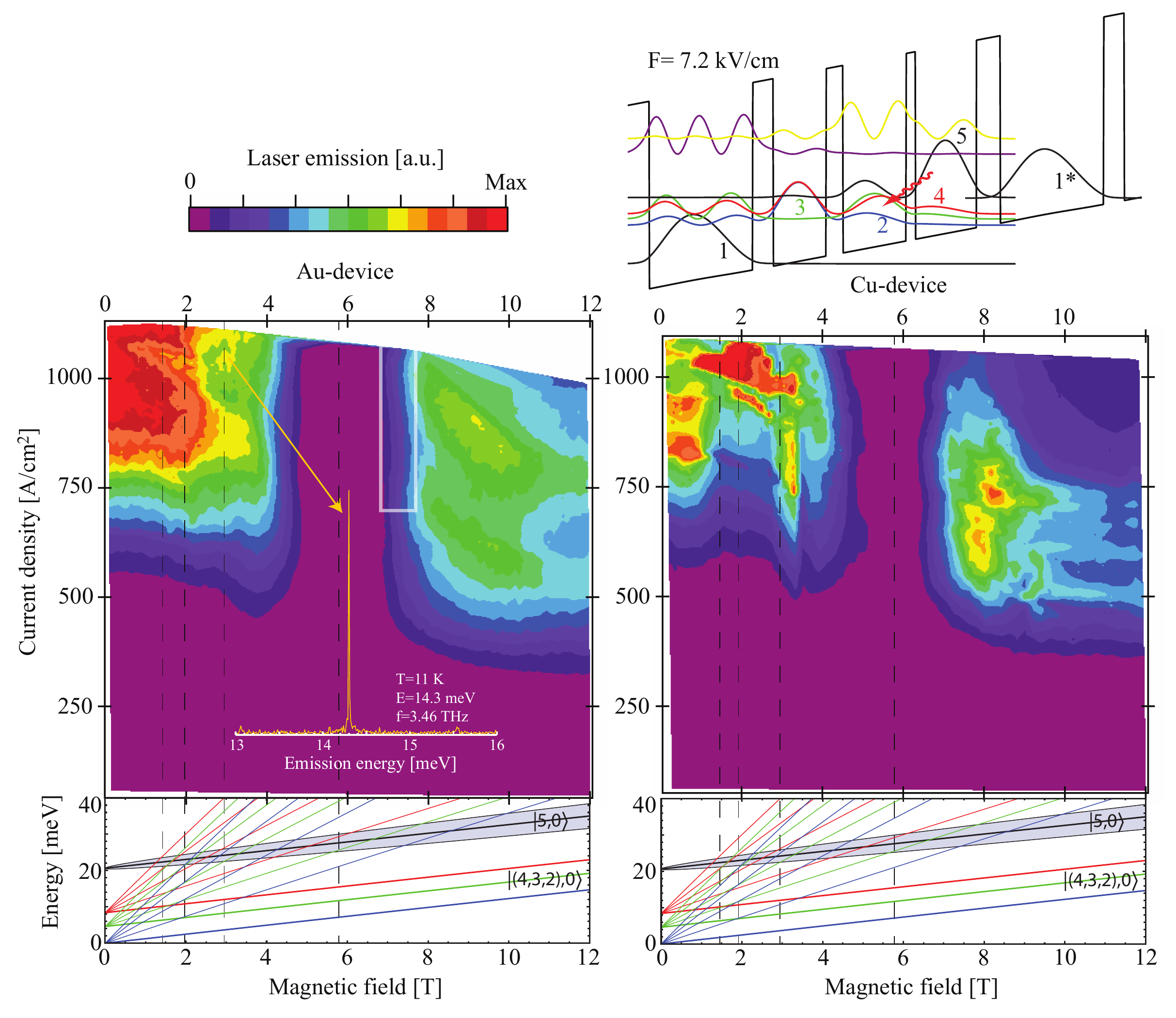}
\caption{QCL emission plotted against magnetic field and injected current density for the Au-device at 4.2 K (left) and the Cu-device at 30 K (right). For clarity the Landau fan of the involved states, calculated with the non-parabolicity-corrected formula, is reported below each map. In the left map, the white rectangle about 7.5 T identifies the region in which lasing at 1.4 THz also occurs, while the spectrum shows the laser emitting at 3.46 THz at B=3 T, T=11 K and J=1077 A/cm$^2$. In the upper-right corner the conduction band diagram of the layer sequence at the alignment is shown.\label{Au_4K}}
\end{figure}

The investigated QCLs use the same gain medium as the best device reported in Ref.\cite{Ohtani2013}, i.e. In$_{0.53}$Ga$_{0.47}$As/Al$_{0.17}$In$_{0.52}$Ga$_{0.31}$As grown lattice matched on InP substrate by Molecular Beam Epitaxy. The heterostructure consists of 150 repetitions of the bound-to-continuum structure with dimensions in \AA $ $ \underline{217}/{\bf 43}/111/{\bf 35}/132/{\bf 20}/128/{\bf 51}, where the barriers are indicated in bold and the wide underlined well serves as injector and is doped with Si with sheet density $n_s$=4.86$\times$10$^{10}$ cm$^{-2}$. The conduction band offset is $\Delta E_{CBO}$=141 meV while the electron effective masses are $m^*_{InGaAs}$=0.043m$_e$ and $m^*_{AlInGaAs}$=0.057m$_e$ for wells and barriers, respectively. A schematic of the conduction band diagram is presented as an inset in Fig.\ref{Au_4K}. The structure is based on a bound-to-continuum design with resonant phonon extraction from Ref.s\cite{Fischer2010, Amanti2009} and relies on an optical transition between state $\Ket{5}$ and states $\Ket{4,3}$ of the $\Ket{4}-\Ket{3}-\Ket{2}$ miniband that gets resonantly depopulated via LO-phonon scattering into state $\Ket{1}$. The structure is aligned for an applied electric field of F= 7.2 kV/cm with the following calculated parameters. The transition energies from the upper state are $E_{54}$=12.3 meV, $E_{53}$=16.0 meV and $E_{52}$=20.6 meV with dipole elements $z_{54}$=6.31 nm, $z_{53}$=2.49 nm and $z_{52}$=0.63 nm, corresponding to oscillator strengths of $f_{54}=12.8$, $f_{53}=2.6$ and  $f_{52}=0.21$. The energy separation and main dipole elements of the lower lying states are $E_{43}$=3.8 meV, $z_{43}$=15.1 nm, $E_{42}$=8.4 meV, $z_{42}$=0.12 nm, $E_{32}$=4.6 meV,  $z_{32}$=11.9 nm, and $E_{21}$=28.8 meV with a depopulation time $\tau_{41-LO}=2.15$ ps at 50 K.

The grown wafer was then processed in metal-metal waveguide configuration with ridges 150$\mu$m wide and about 1.5mm long, some with a Au-Au waveguides (Ti/Au 5/500nm), some with a Cu-Cu waveguides (Ti/Cu 5/500nm) in order to reduce the waveguide losses\cite{Williams2005,Belkin2008}. In the following, the two different device types will be denoted ``Au-device'' and ``Cu-device''. The Cu-device was first characterised at B=0 T and it reached a maximum operating temperature of 148 K, showing an 18 K-improvement with respect to the Au-device.\cite{Ohtani2013}


The measured devices were driven in current with a pulser in a macro-micro pulses configuration. The source signal is constituted by a burst of 500 to 1000 micro-pulses with widths between 95 and 345 ns (duty cycle between 0.3 and 2\%). The signal is then square-wave-modulated at 30 Hz to match the liquid-Helium-cooled bolometer response. In the following, the specific parameters will be pointed out along to each measurement.

For the present study, the J-V characteristics of the QCLs were recorded along with the laser emitted intensity (L) as a function of the magnetic field applied along the growth axis within the temperature range 4-200 K. In Fig.\ref{Au_4K} the B-J-L map for the Au-device is presented on the left (1000 pulses per macro-pulse with width 95 ns, T=4.2 K). Below the map, the Landau level fans for the involved upper state and lower miniband are plotted for the first few orders, according to the non-parabolicity-corrected formula\cite{Ekenberg1989,Sirtori1994} $E(B,n)=\frac{1}{2}(E(0)-E_G)+\frac{1}{2}\sqrt{(E(0)-E_G)^2+4E_G\left(E(0)+\left(n+\frac{1}{2}\right)\frac{\hbar e B}{m^*(0)}\right)}$ where $E(0)$ is the energy of the state at B=0 T, $E_G=816$ meV is the gap energy, $n$ the Landau level index, $\hbar$ the reduced Planck constant and $m^*(0)$ is the effective mass at B=0 T. The bold lines mark the Landau index n=0 of the upper state $\Ket{5,0}$ and of the lower-lying miniband $\Ket{(4,3,2),0}$. For the level $\Ket{5,0}$ broadening is taken into account according to the formula $\Gamma=\Gamma_0\sqrt{B}$ assuming $\Gamma_0$=1 meV  at B=0T.\cite{ANDO1974, Alton2003} Equivalently, one could instead consider the un-broadened state $\Ket{5,0}$ interacting with a continuum of states delimited by $\Ket{4,n}$ and $\Ket{3,n}$. Thinner lines constitute the higher index Landau levels while dashed lines are indicating the B-field position of their crossings with $\Ket{5,0}$. One has also to note that the lower effective mass of InGaAs with respect to GaAs allows one to reach a 1.6-times ($m^*_{GaAs}/m^*_{InGaAs}$=67/43) higher confinement for the same magnetic fields. 

\begin{figure}[hbtp]
\centering
\includegraphics[height=.8\textheight]{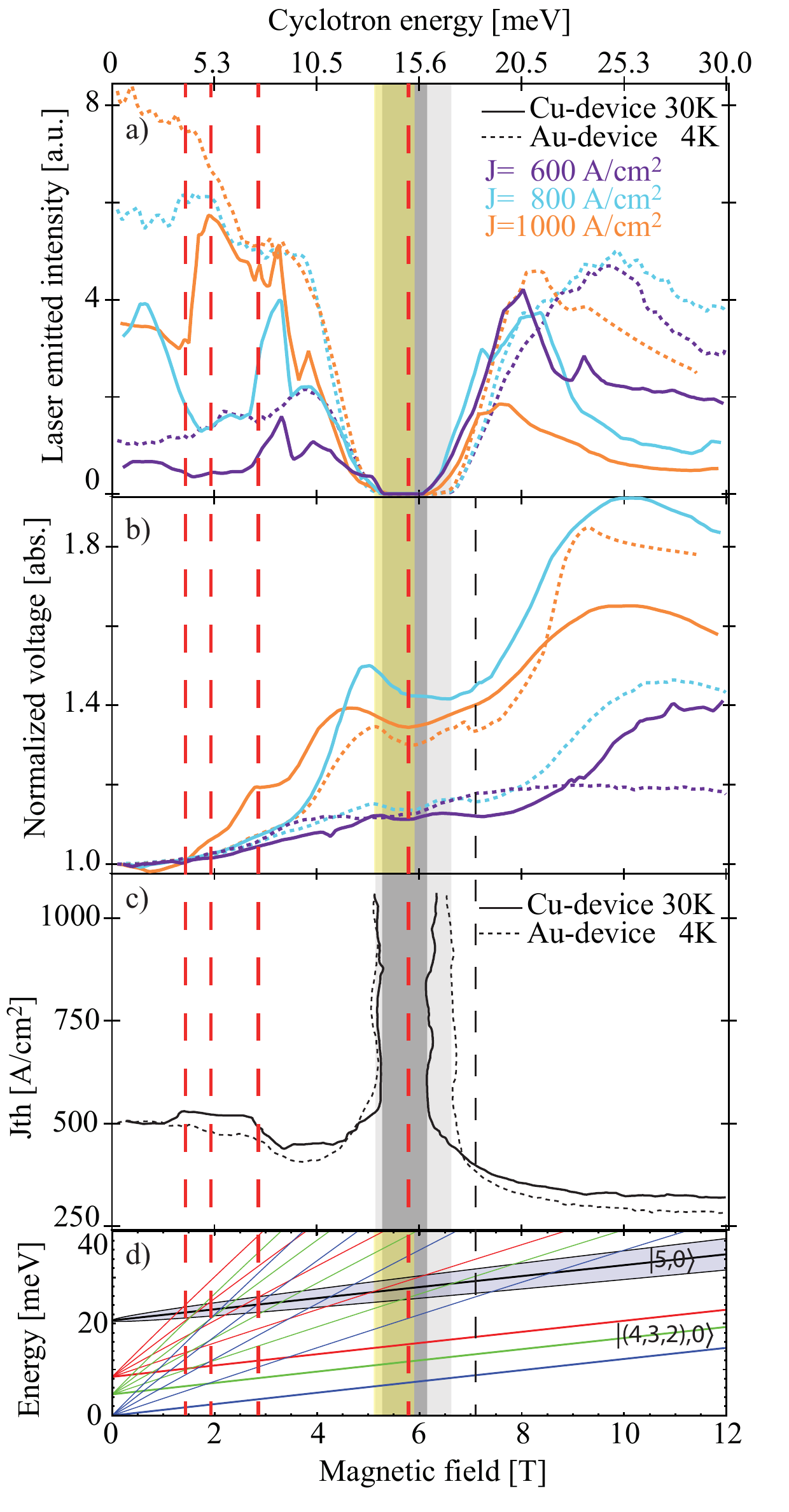}
\caption{a)Laser emission, b)Normalized voltage V(B)/V(B=0 T), c)Threshold current density, d)Landau fan chart of the involved states. All curves are plotted against magnetic field for the Au- and Cu-devices, displayed in dashed and full lines, respectively. The curves of panel a) and b) are taken at fixed current densities: 600, 800, 1000 A/cm$^2$ displayed in violet, light blue and orange, respectively. Additional explanations can be found in the main text.\label{LB}}
\end{figure}

Upon inspection of the map for the Au-device in Fig.\ref{Au_4K}, it is evident that the laser emits in two B-field regions (first for B$<$5.1 T, second for B$>$6.6 T) separated by a gap extending for about 1.5 T, as better visible from Fig.\ref{LB}(a).  As already observed in previous investigations\cite{ScalariLPR2009,Scalari2010, Wade2009,Maero2013} and supported by the transport data discussed below, when the lasing transition $\Ket{5}\rightarrow \Ket{4,3}$ is resonant with the cyclotron energy elastic scattering opens a relaxation channel and the population inversion decreases. As a consequence the gain decreases and eventually falls below the losses' value ceasing the lasing action. This is the case at 5.8 T (15.1 meV) where $\Ket{(4,3),1}$ cross $\Ket{5,0}$. 
On the right-hand side of Fig.\ref{Au_4K} the B-J-L map of the Cu-device is shown (1000 pulses per macro-pulse with width 345 ns, T=30 K). One can immediately recognise that the central gap is narrower, extending for 1 T vs 1.5 T. This directly points at a lower losses' level for the Cu-Cu waveguide with respect to the Au-Au. 

Several oscillations in the threshold current density and consequent variations in the emitted light are present in the colour maps, especially for the low B-field region. These features are particularly strong for the Cu-device. Most of them are related to the crossings of the higher index Landau levels of states $\Ket{4}$ and $\Ket{3}$ with the upper lasing state $\Ket{5}$, therefore modulating the level of gain and population inversion. 

\begin{figure}[hbtp]
\centering
\includegraphics[width=\linewidth]{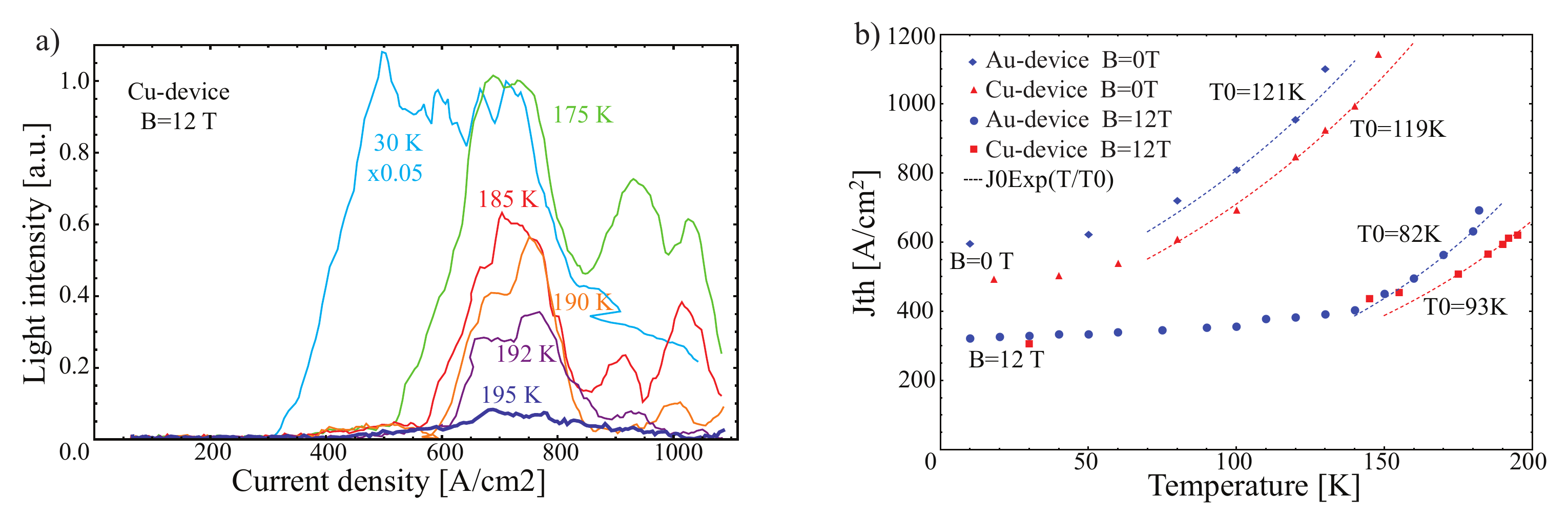}
\caption{a) L-J curves for the Cu-device at B=12 T for temperatures of 30 K (light blue), 175 K (green), 185 K (red), 190 K (orange), 192 K (violet) and 195 K (blue, thick). b) J$_{th}$ vs Temperature for both devices at B=0 T and B=12 T.\label{LIVs}}
\end{figure}
The lasing performance of the QCLs were studied at different magnetic fields and temperatures.  The first benefit of the in-plane carrier confinement is the improvement of the maximum operating temperature of the lasers: the Cu-device (Au-device) raised from 148 K (130 K) to 195 K (182 K) when applying a magnetic field of intensity 12 T. The J-L curves of the Cu-device at selected temperatures at 12 T are presented in Fig. \ref{LIVs}a) and reveal the laser emission performance till the maximum emission temperature of 195 K (thick blue line).

The temperature behaviour of the threshold current density (J$_{th}$) is shown in Fig.\ref{LIVs}b) at B=0 T and B=12 T for both devices. The 0 T-series is extracted from measurements performed in an external flow cryostat coupled to a bolometer (the difference in the values of J$_{th}$ at low temperatures between the maps and the points here reported might stem from a different reading of the temperature sensor). Comparing the performances at B=0 T, one can see how the Cu-device has a lower threshold current density in the whole temperature range that allows it to reach a higher maximum temperature, despite of the very similar T0. 

J$_{th}$ is related to the losses as\cite{JeromeBook} 
\begin{equation}
J_{th}=\frac{e\alpha_{tot}}{g_c\tau_{eff}}\quad \textrm{with}\quad \alpha_{tot}=\alpha_m+\alpha_w+g_{ISB}n_s
\label{Jth_alpha}
\end{equation}
where $e$ is the electron charge, $\alpha_{tot,m,w}$ the total/mirror/waveguide losses. $g_{c}$ is the gain cross section defined as $g_{c}=\Gamma \frac{2\pi e^2z^2}{\epsilon_0 n_{refr}L_p\lambda\gamma}$ with $\Gamma$ being the overlap factor with the waveguide, $z$ the dipole of the transition, $\epsilon_0$ the vacuum permittivity, $n_{refr}$ the active region refractive index, $L_p$ the period length and $\lambda$ and $\gamma$ the transition wave length and broadening. $g_{ISB}$ is the intersubband (ISB) cross section due to reabsorption by free carriers in the heterostructure. $g_{ISB}n_s$ gives then an estimate of the free carrier loss. The effective upper state lifetime, $\tau_{eff}=\tau_{up}(1-\tau_{dn}/\tau_{up\rightarrow dn})$, accounts for the non-zero lower state lifetime $\tau_{dn}$ and for the finite laser transition scattering rate $\tau_{up\rightarrow dn}$. Since all devices are processed from the same heterostructure, we can assume that, at a fixed magnetic field and temperature, $g_c$, $\tau_{eff}$, $g_{ISB}$ and $n_s$ are the same. Therefore, from eq.\eqref{Jth_alpha} the ratio of the J$_{th}$ is proportional to the ratio of the total losses. 

On the whole temperature range, the 0 T-series maintain approximately the same J$_{th}$ ratio Jth$^{Cu}_{0T}$/Jth$^{Au}_{0T}$=$\alpha_{tot}^{Cu}$/$\alpha_{tot}^{Au}$=0.85$\pm$0.03, showing a 15\% loss reduction. After Ohtani\cite{Ohtani2013}, the mirror and waveguide losses of the Au-device at 10 K amount to $\alpha_m^{Au}+\alpha_w^{Au}$=(7.2$\pm$2.9)cm$^{-1}$ with total losses $\alpha_{tot}^{Au}=(17.5\pm 4)$cm$^{-1}$ (we can assume that all devices in Ref.\cite{Ohtani2013} have similar free carrier losses since the doping levels and the structures are very close to each other). This finally allows us to derive the Cu-device total losses $\alpha_{tot}^{Cu}$=(14.9$\pm$3)cm$^{-1}$. Since the two device types differ only in the waveguide materials and because of Eq.\eqref{Jth_alpha}, one can derive $\alpha_m^{Cu}+\alpha_w^{Cu}$=(4.6$\pm$6)cm$^{-1}$ that corresponds to a reduction of one third of the cavity losses with respect to the Au-device, consistently with what reported in the literature.\cite{Belkin2008} The large error is the result of a pessimistic error estimation via standard correlated error propagation and we want to stress that the possible range for the mirror and waveguide losses is meant to be bound to positive values, given Cu-total losses of about 14.9 cm$^{-1}$.

When applying the magnetic field, the low-temperature J$_{th}$ gets reduced by almost a factor 2 and the lasers' performances in temperature change. Up to about 160 K, the J$_{th}$s of both device types are very close and increase almost linearly in T, showing a very weak temperature dependence. This fact indicates that, in this range, the J$_{th}$ is not set any more by the losses: now J$_{th}$ is dominated by the transparency current needed to keep the structure aligned and the laser performance is voltage-limited. Such an effect has already been seen in studies of other THz-QCLs.\cite{Benz2007} It is only from 160 K that the Cu-device outperforms the Au-device. At the highest temperatures, the J$_{th}$ ratio for the two devices is consistent with the one without applied B-field, namely Jth$^{Cu}_{12T}$/Jth$^{Au}_{12T}$=0.87$\pm$0.05, indicating once more that the Cu-losses are lower than the Au ones. 

\begin{figure}[hbtp]
\centering
\includegraphics[width=.8\linewidth]{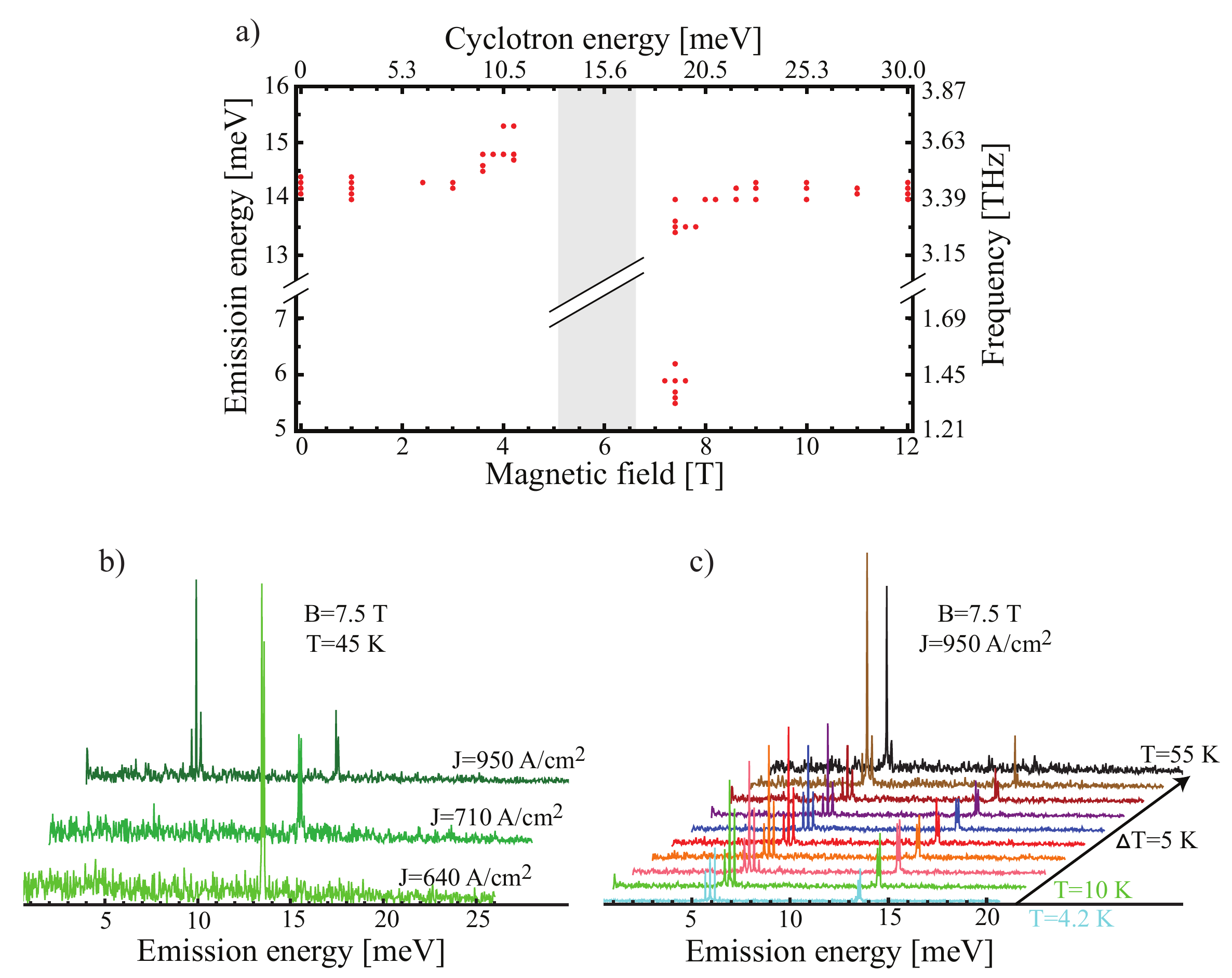}
\caption{a) Emission energy chart for the Au-devices plotted against magnetic field. b) and c) Spectra from Au-devices showing the behaviour of the low-energy emission in J and T. (500 pulses per macro-pulse with width 104 ns)\label{emission}}
\end{figure}

A study of the emission spectra was performed on the Au-device. Spectra were taken at several points of operation of the lasers for different values of magnetic field, injected current and temperature. When considering the strongest emission line of each recorded spectrum vs B-field, the emission energy chart of Fig.\ref{emission}a) can be compiled: the lasers mostly emit at 14.3 meV. Approaching the non-lasing region, the emission energy experiences a blue- (red-) shift coming from lower (higher) magnetic field values. This behaviour can be attributed to the Stark shift originated by the increasing voltage applied to the laser needed to reach the threshold current density while competing with the increasing non-radiative losses. This is also supported by Fig.\ref{LB}b) where the normalized voltage at fixed J is shown. Here two voltage shoulders are present on each side of the gap region (in grey).

Additionally, at magnetic fields about 7.5 T (rectangle in Fig.\ref{Au_4K}) the device is found to lase also in the range 5.5-6.3 meV (1.33-1.52 THz). Comparing the emission energy with the calculated levels and dipoles, it is consistent with a transition from state $\Ket{3}$ to the lower one $\Ket{2}$ ($E_{32}$=4.6 meV and $f_{32}$=17.1). We investigated this emission in temperature and current density: the resulting spectra are collected in Fig.\ref{emission} b) and c), respectively. The first series shows the current density dependence: the low-energy emission is not present immediately after threshold, but appears with increasing injected current, at the expenses of the high-energy one. This is most likely due to the fact that, close to the non-lasing gap, the resonance between the photon and the cyclotron still depopulates state $\Ket{5}$ in favour of $\Ket{4,3}$  thus accumulating carriers in $\Ket{3}$ that is now able to achieve population inversion and lase onto state $\Ket{2}$. The same effect was found in similar structures\cite{Scalari2010}. Then, at fixed J=950 A/cm$^2$, the temperature dependence is presented in the second series: the low-energy emission is present at 4.2 K and is of comparable intensity to the higher energy one (cyan spectrum). Increasing the temperature, the higher-energy emission, now at 13.5 meV, slightly decreases in intensity and coexists till 50 K. At 55 K the low-energy emission is the only one present and is much stronger than at lower temperatures. Further increment of the temperature brings back the system to the high-energy emission only.

In Fig.\ref{LB} different quantities useful to compare the laser behaviour in magnetic field are sequentially plotted as a function of the magnetic field as continuous and dashed lines for the Au- and Cu-device, respectively. Panel a) compares the emitted intensity at fixed injected current densities of 600, 800 and 1000 A/cm$^2$ (horizontal cuts of the maps in Fig.\ref{Au_4K}). 
Panel b) shows the normalized voltage at constant current density, while panel c) compares the threshold current densities. Finally panel d) reports the energy position of the states including the calculated Landau fan. 
 
The curves in panel a) first clearly show the difference in the extension of the non-lasing region between the two devices. The two regions are indicated in light and dark grey throughout Fig.\ref{LB} and, as seen from panel c), they are maintained over the whole explored current density range. As already noted, they take place about the main crossing of the upper laser level with the first Landau level of the lower ones at 15.1 meV. This crossing is determined in the V(B) curves in panel b) by the main broad central minimum present in most curves and identifies the energy of the electronic transition between the upper and lower laser levels. In fact when the resonance condition is met, a non-radiative channel opens for electrons to decay faster than the radiative transition, requiring less applied bias to sustain the same current density. Considering Eq.\eqref{Jth_alpha}, this means a strong decrease of $\tau_{up}$ resulting in a big increase in J$_{th}$. 

In the lower B-field range there are several features both in emission and transport. They are much stronger for the Cu-device, and take place close to the main crossing of $\Ket{5,0}$ with $\Ket{(4,3),n}$. In the high B-field region, the laser is in the strong confinement regime\cite{Scalari2007} where no relevant crossing takes place and the decrease of the emission with higher magnetic field is due to the progressive misalignment of the states with increasing magneto-resistance.

The range of emission of the QCLs between 13.4 and 15.3 meV is reported throughout Fig.\ref{LB} as a yellow band, corresponding via the cyclotron energy to the range 5.1-5.9 T. The extreme energy values of the emission have been already discussed in the previous paragraphs. When instead considering the `asymptotic' values for low and high B-fields, i.e. 14.3 meV, it is clear that the optical transition is shifted with respect to the electronic one derived from transport to be at 15.1 meV. Such discrepancy has already been seen in THz-QCLs\cite{Ulrich2000,Scalari2005} and stems from the fact that the matrix elements ruling the transition from the upper to the lower subband have different expressions for the optical transition\cite{Helm2000} and in the case of electron-electron scattering between the respective Landau levels\cite{Kempa2002}. These then result different energies when summed over the different states of the miniband. Alternatively, several many body effects have been pointed out as responsible for similar shifts in highly doped inter-subband systems, the main of which is the depolarization shift\cite{Helm2000,Kisin1998}. Its amplitude calculated for the present system with a refractive index $\epsilon_r$=13.32, a quantum well width $a=22$ nm and assuming\cite{Amanti2009} $n_{up}=10\% n_s=4.86\times 10^{9}$ cm$^{-2}$, amounts to 10\% of the measured energy difference. The depolarization shift can have therefore a sizeable influence on the transition energy but the present carrier density is too low to account for the full shift.

Noteworthy is then the fact that in Fig.\ref{LB}b) there is a feature showing a minimum at about 7.1 T that appears in all curves with different strength and is especially evident for the Au-device (thin dashed line). Since it is close to the magnetic field value for which lasing at low energy is detected, we attribute this minimum to a joint effect of the crossing of $\Ket{3,1}$ with $\Ket{5,0}$ and photon assisted transport from state $\Ket{3}$ to $\Ket{2}$ while lasing at about 5.9 meV.

Finally, it is interesting to compare the different material systems which have been studied with magneto-spectroscopy. The relevant quantities are summarized in Table \ref{tab}. All three lasers emit at similar energies and strikingly their temperature performances improve of the same amount (+48 K) independently of the material system. Three energies are relevant for the present comparison: the photon energy  E$_{phot}$, the LO-phonon energy E$_{LO}$ and the cyclotron energy in the wells at the magnetic field for which the maximum operating temperatures were recorded E$_c($B$|_{T_{max}}$). When taking the ratios between them, one realizes that all maximum temperatures were recorded with the laser in the ultra-strong confinement regime\cite{Scalari2007,Wade2009} when E$_c($B$|_{T_{max}}$) is greater than the photon energy and close to the phonon energy . This seems to underline some common operation regime/limitations and in fact it is consistent with the fact that all designs are based on resonant phonon depopulation that gets enhanced when the cyclotron is as well resonant. This might be interpreted as the current lasers having all about 50 K to gain in performance from the optimal depopulation of the lower level. On the other hand, such improvement would not be enough and  optimization of the other parameters, like scattering mechanisms, transition type and band structure, is fundamental for THz-QCL aiming at operating close to room temperature.

\begin{table}[htdp]
\footnotesize
\caption{Material system comparison for the THz-QCLs investigated also in strong magnetic field: for each system the effective masses $m^*$ for barrier and well are reported along with conduction band offset $\Delta E_{CBO}$, metal constituting the waveguide (Wvg.), maximum operating temperature with and without magnetic field with their difference $\Delta T_{max}$= T$_{max}$(B)-T$_{max}$. Photon and LO-phonon energies (E$_{phot}$, E$_{LO}$) are also reported together with the ratio of the cyclotron energy calculated at the magnetic field where the maximum temperature was measured $E_c(B|_{T_{max}})$ with the previous ones. In order to have a fair comparison, all cyclotron energies in this table were computed in the constant-mass approximation.}
\resizebox{\textwidth}{!}{
\begin{tabular}{c|c|c|c|c|c|c|c|c|c|c}
\multirow{2}{*}{Material system} &  m*$_{well}$/$_{barrier}$ & $\Delta E_{CBO}$ &  Wvg. & T$_{max}$ & T$_{max}$(B) &  $\Delta T_{max}$& E$_{phot}$ &  E$_{LO}$ &  \multirow{2}{*}{$\frac{E_c(B|_{T_{max}})}{E_{phot}}$}&  \multirow{2}{*}{$\frac{E_c(B|_{T_{max}})}{E_{LO}}$} \\
&(m$_e$)&  (meV)  &  Metal & (K)& (K)&  (K)& (meV) & (meV)& &\\ \hline \hline
GaAs/Al$_{0.15}$Ga$_{0.85}$As & 0.067/0.080 & 150 & Cu &178\cite{Belkin2008} & 225\cite{Wade2009} (19.3 T) & 47 & 13.0 & 36&  2.6 &  0.93 \\  \hline
In$_{0.53}$Ga$_{0.47}$As/GaAs$_{0.51}$Sb$_{0.49}$ & 0.043/0.045 &360  & Au&142\cite{Deutsch2012} & 190\cite{Maero2013} (11 T) & 48& 14.8 &33  & 2.0 & 0.90\\  \hline
 \multirow{2}{*}{In$_{0.53}$Ga$_{0.47}$As/Al$_{0.17}$In$_{0.52}$Ga$_{0.31}$As} & \multirow{2}{*}{0.043/0.057} &  \multirow{2}{*}{141}  &   Cu  & 148 &195 (12 T) & 47 & \multirow{2}{*}{14.3}  &\multirow{2}{*}{33} &   \multirow{2}{*}{2.3} &  \multirow{2}{*}{0.98}\\ 
&&&Au&130\cite{Ohtani2013} & 182 (12 T) &52 &&&&
\end{tabular}
}
\label{tab}
\end{table}%

In conclusion, we presented a study of InGaAs/AlInGaAs THz quantum cascade lasers in magnetic field and with two different waveguide metals. We could show that the active region processed in a Cu-Cu waveguide lases at about 14.3 meV (and for a small interval of magnetic field and temperature, also at 5.9 meV) up to 148 K without magnetic field and up to 195 K with B=12 T. The maximum operating temperatures are 18 K and 13 K higher than the ones for Au-Au waveguide devices, for the case without and with magnetic field, respectively. This confirms the better temperature performance of the devices with Copper waveguide. Finally, the comparison of the studies done on the different material systems shows that by an optimization of the extraction mechanism one can gain up to 50 K of operating temperature range in pulsed regime.
\begin{acknowledgments}
The authors acknowledge support by SNF and ETH Zurich and want to thank C. Bonzon, M. R\"osch and D. Turcinkova for help and discussion.
\end{acknowledgments}

%

\end{document}